\documentclass[%
reprint, 
superscriptaddress,
nofootinbib,
 amsmath,amssymb,
 aps, physrev,
]{revtex4-2}

\usepackage{graphicx}
\usepackage[percent]{overpic}
\usepackage[caption=false]{subfig}
\usepackage{physics}
\usepackage[dvipsnames,table]{xcolor}
\usepackage[colorlinks=true,linkcolor=RoyalBlue,citecolor=Brown,urlcolor=Brown,filecolor=Brown]{hyperref}
\usepackage[normalem]{ulem}

\usepackage{booktabs,todonotes,upgreek}

\makeatletter
\def\CT@@do@color{%
  \global\let\CT@do@color\relax
        \@tempdima\wd\z@
        \advance\@tempdima\@tempdimb
        \advance\@tempdima\@tempdimc
\advance\@tempdimb\tabcolsep
\advance\@tempdimc\tabcolsep
\advance\@tempdima2\tabcolsep
        \kern-\@tempdimb
        \leaders\vrule
                \hskip\@tempdima\@plus  1fill
        \kern-\@tempdimc
        \hskip-\wd\z@ \@plus -1fill }
\makeatother

\begin{document}

\title{\textbf{Experimental Realization of a Rotating Radio-Frequency Ion Trap for Precision Metrology} 
}
\author{Sun Yool Park}
\email{sun.park@colorado.edu}
\affiliation{JILA, NIST and University of Colorado, and Department of Physics, University of Colorado, Boulder CO 80309, USA.}

\author{Anzhou Wang}
\affiliation{JILA, NIST and University of Colorado, and Department of Physics, University of Colorado, Boulder CO 80309, USA.}

\author{Kia Boon Ng}
\affiliation{TRIUMF, Vancouver, British Columbia, V6T 2A3, Canada}

\author{Patricia Hector Hernandez}
\affiliation{JILA, NIST and University of Colorado, and Department of Physics, University of Colorado, Boulder CO 80309, USA.}

\author{Addison Hartman}
\affiliation{JILA, NIST and University of Colorado, and Department of Physics, University of Colorado, Boulder CO 80309, USA.}

\author{Tuan Anh Nguyen}
\affiliation{JILA, NIST and University of Colorado, and Department of Physics, University of Colorado, Boulder CO 80309, USA.}

\author{Rohan Kompella}
\affiliation{JILA, NIST and University of Colorado, and Department of Physics, University of Colorado, Boulder CO 80309, USA.}

\author{Michail Athanasakis-Kaklamanakis}
\affiliation{JILA, NIST and University of Colorado, and Department of Physics, University of Colorado, Boulder CO 80309, USA.}

\author{Jun Ye}
\affiliation{JILA, NIST and University of Colorado, and Department of Physics, University of Colorado, Boulder CO 80309, USA.}

\author{Eric A. Cornell}
\affiliation{JILA, NIST and University of Colorado, and Department of Physics, University of Colorado, Boulder CO 80309, USA.}

\begin{abstract}
We discuss the experimental realization of the rotating radio-frequency (rrf) trap, proposed by Hasegawa and Bollinger [Phys. Rev. A \textbf{72}, 043403 (2005)]. Compared to a traditional linear rf (lrf) Paul trap, the rrf trap is a closer analogy to the popular mechanical lecture demonstration for a Paul trap. In an ion trap with reslistic, non-ideal electrode geometry, the rrf trap averages over angular variations in the effective potential. This averaging provides more uniform confinement and reduces ion loss at equal confinement strength compared with the lrf trap. This feature makes the rrf trap configuration advantageous for precision metrology application, such as electron electric dipole moment (eEDM) measurements.

\end{abstract}

\maketitle

\section{Introduction}
\begin{figure}[t] 
    \centering
    \subfloat[]{
    \includegraphics[width=1 \linewidth]{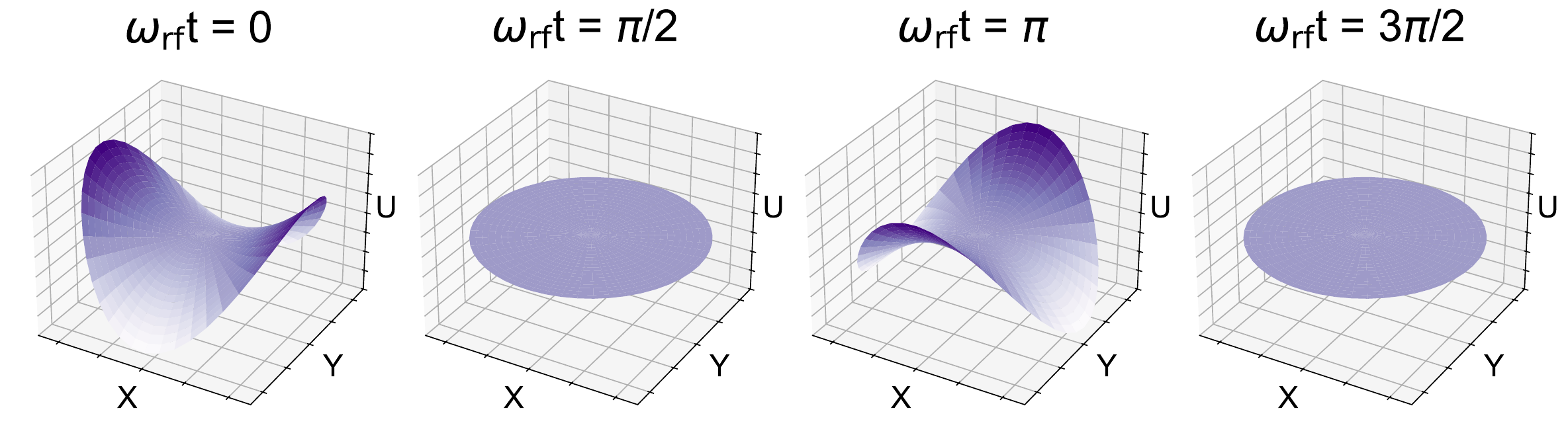}
    \label{fig:lrf_landscape}
    }\hfill
    \subfloat[]{
    \includegraphics[width=1 \linewidth]{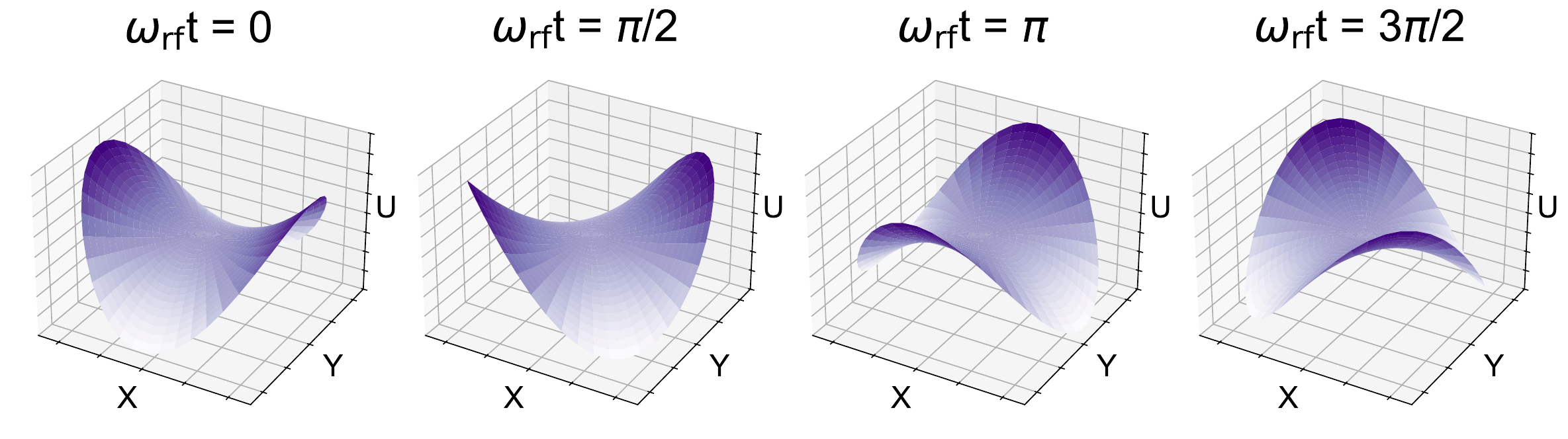}
    \label{fig:rrf_landscape}
    }\hfill
    \caption{The landscape of the oscillating component of the potential of the (a) linear rf ion trap and the (b) ``mechanical" Paul trap at different moments in time $t$. (a) Linear rf ion trap. The potential is confining (anticonfining) along the $x$-axis ($y$-axis) at $\omega_\mathrm{rf} t=0$. A quarter period later, the potential landscape is flat. Half a period later, the potential flips its sign. The potential flaps along the $x$ and $y$ axes. (b) Mechanical Paul trap. The potential rotates (in this case, about the $+\hat{z}$ direction). The same pattern appears in the rrf ion trap.}
    \label{fig:potential_landscape}
\end{figure}

Radio-frequency (rf) ion traps, commonly referred to as Paul traps, allow for confinement in all three directions. Since three-dimensional ion confinement using purely electrostatic fields is not a solution of Laplace's equation, linear rf (lrf) ion traps utilize alternating confining and anticonfining potentials oscillating along the $x$ and $y$ axes as shown in Fig.~\ref{fig:potential_landscape}a. The rapidly oscillating fields drive micromotion, and this micromotion in the field gradient results in an effective secular confining potential.
Since their introduction in the 1950s by Wolfgang Paul \cite{Paul1953}, lrf ion traps have been used in many applications, including quantum computation \cite{CiracZoller1995, Monroe1995, Haffner2008, BlattWineland2008, Main2025DistributedDQC, Liu2025CertifiedRandomness}, atomic clocks \cite{Diddams2001, Ludlow2015, Marshall2025PRL_AlClock, Lindvall2025PRApplied_SrClock, Cheung2025PRL_Ni12Clock}, and physical chemistry \cite{Willitsch2012, Heazlewood2019, KrohnLewandowski2024}.

There is a familiar mechanical Paul trap lecture demonstration (videos available on the internet, for instance here in Ref.~\cite{HarvardNatSciDemosRotatingSaddleVideo}) based on a rotating, saddle-shaped surface. 
An object constrained to the surface by gravity and surface forces experiences a potential that rotates with time as in  Fig.~\ref{fig:potential_landscape}b. While this is not exactly how most rf ion traps work, the analogy provides useful physical intuition. 

About 20 years ago, Hasegawa and Bollinger in Ref.~\cite{Bollinger_2005} proposed an rf ion trap based on a rotating potential. In conventional lrf ion traps, the confining and anticonfining potentials flap back and forth at the rf frequency, whereas in the rotating rf (rrf) ion trap, the potential itself rotates at the rf frequency, more analogous to the mechanical lecture demonstration. 
Prior to this work, no experimental realization of an rrf ion trap has been reported in the literature. This could be because an rrf ion trap is experimentally more complex, requiring multiple phases of rf voltages on multiple electrodes. 

In this paper, we revisit this proposal and present an experimental realization of the rrf ion trap, verifying its fundamental dynamics and demonstrating the theoretical predictions outlined in Ref.~\cite{Bollinger_2005}. Furthermore, we demonstrate practical implications of the rrf ion trap for precision metrology. Specifically, we show how the rrf ion trap configuration resolves inherent trap-depth anisotropies in eight-rod trap geometries used in precision spectroscopy on trapped molecular ions. 

Our group at JILA recently set an upper bound on the electron's electric dipole moment (eEDM) of $|d_e| < 4.1 \times 10^{-30}$ $e$~cm using molecular ions in a linear rf trap \cite{Roussy_Science}, and we are attempting to further extend the precision \cite{ng2023thesis, Cairncross2019}. The sensitivity of the eEDM measurement is proportional to $\sqrt{N} \tau$ where $N$ is the number of detected ions and $\tau$ is the coherence time. 
Therefore, to maximize sensitivity, we aim to trap as many ions as possible while suppressing unwanted ion--ion interactions by adiabatically reducing the secular frequency, which lowers the ensemble temperature and density. Throughout this paper, we distinguish between the \textit{tightness} of ion confinement proportional to the secular frequency at the trap center and the \textit{effective depth} of the trap set by the total energy required to escape the trap via the lowest saddle point. While loose confinement can be advantageous for reducing interaction-driven decoherence, it also makes the ions more vulnerable to confining-potential inhomogeneity: the effective trap depth becomes shallower, which increases ion loss. The rrf trap mitigates this tradeoff by averaging over certain static inhomogeneities and anharmonicities, yielding a deeper effective trap depth for the same nominal confinement, as will be discussed later in this paper.

\section{Theory and Experimental Realization of the rrf trap} \label{sec:idealized_rf}
\subsection{Idealized Trap Potentials}
In an ideal lrf ion trap, an ion experiences a dc axial potential $\Phi_z$ and an oscillating radial quadrupole potential $\Phi_\mathrm{lrf}$ given in cylindrical coordinates $(r, \theta, z)$ by
\begin{equation} \label{eq:lrf_potential}
\begin{split}
\Phi_z & = - \frac{V_\mathrm{dc}}{r_\mathrm{a}^2} (r^2-2z^2),\\
\Phi_\mathrm{lrf} & = \frac{V_\mathrm{rf}}{r_\mathrm{r}^2}r^2\cos{(2\theta)}\cos{(2\omega t)},
\end{split}
\end{equation}
where $V_\mathrm{dc}$ and $V_\mathrm{rf}$ are the dc and the rf voltages applied, $r_\mathrm{a}$ and $r_\mathrm{r}$ are effective axial and radial dimensions of the end caps and radial electrodes, and $2\omega=\omega_\mathrm{rf}$ is the rf frequency \cite{Bollinger_2005}. By contrast, in an rrf trap, the rf radial potential rotates with time as \cite{Bollinger_2005}
\begin{equation} \label{eq:rrf_potential}
    \Phi_{\mathrm{rrf}} = \frac{V_\mathrm{rf}}{r_\mathrm{r}^2}r^2\cos{[2(\theta+\omega t)]}.
\end{equation}
There are two dimensionless Mathieu parameters in rf traps: $q$ and $d$ are defined as
\begin{equation}
\begin{split}
    q & = \frac{4 q_0 V_\mathrm{rf}}{m_0{\omega_\mathrm{rf}}^2{r_\mathrm{r}}^2}, \\
    d  & = \frac{4 q_0 V_\mathrm{dc}}{m_0{\omega_\mathrm{rf}}^2{r_\mathrm{a}}^2}, 
\end{split}
\end{equation}
where $q_0$ is the charge of the ion and $m_0$ is the mass of the ion. The $q$ parameter describes the strength of the rf electric field while $d$ describes the applied dc quadrupole field.

Ion motion in an idealized lrf trap has been discussed extensively (see, e.g., Refs.~\cite{Leibfried2003, Major2005ChargedParticleTraps}).
For an rrf ion trap, Hasegawa and Bollinger solved the equations of motion in the rotating, time-varying potential in Ref.~\cite{Bollinger_2005}. 
In both lrf and rrf traps in the small-$q$ limit, the transverse motion separates into a fast micromotion at $\omega_\mathrm{rf}$ and a slower secular motion.
In an ideal lrf potential of Eq.~\eqref{eq:lrf_potential}, the two transverse secular modes are degenerate, with the eigenfrequency 
$\omega_0 = (\omega_\mathrm{rf}/2)\sqrt{q^2-d}$. 
One convenient basis for these degenerate modes is clockwise (CW) and counterclockwise (CCW) circular motion. In practice, however, static transverse stray fields (often well approximated by an $x^2-y^2$ quadrupole term) lift this degeneracy and select two orthogonal linear normal modes, typically aligned close to the $x$- or $y$-axis, resulting in two distinct secular frequencies $\omega_x \neq \omega_y$. In this case, the mode axes are fixed in the laboratory frame, and generic excitations lead to beating between the two linear modes.

In the rrf trap, the degeneracy is lifted intrinsically by the rotating rf drive. For a finite value of $q$, the two transverse secular eigenmodes are split even in the absence of static transverse fields. To leading order in $q$, the CW and CCW eigenfrequencies are \cite{Bollinger_2005}
\begin{equation}\label{eq:rrf_freq}
\omega_{\pm} \simeq \frac{\omega_\mathrm{rf}}{2}\!\left(\sqrt{q^2-d}\ \pm\ \frac{q^2}{2}\right).
\end{equation}
Because the rrf eigenmodes remain CW and CCW, a linearly polarized excitation along the $x$- or $y$-axis can be viewed as an equal-amplitude superposition of these eigenmodes. 
The resulting transverse motion stays linear, oscillating at $\bar{\omega}_{\mathrm{sec}}=(\omega_++\omega_-)/{2}$, but its axis rotates slowly in the $xy$-plane at the precession rate $\omega_{\mathrm{prec}}=(\omega_+-\omega_-)/{2}$.
% $\simeq \frac{\omega_\mathrm{rf}}{2}\,\frac{q^2}{2}$. 
Exciting instead a pure CW or CCW mode produces a circular secular trajectory for which no oscillation axis is defined, and thus there is no meaningful notion of ``precession.''

\subsection{The JILA Eight-Rod Trap}
\begin{figure}[t]
\centering
\subfloat[][Axial view of the electrodes.\label{fig:setup_a}]
{\includegraphics[width=0.30\columnwidth]{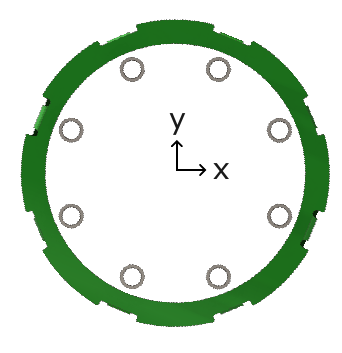}}
\hfill
\subfloat[][Side view of the electrodes.\label{fig:setup_b}]
{\includegraphics[width=0.67\columnwidth]{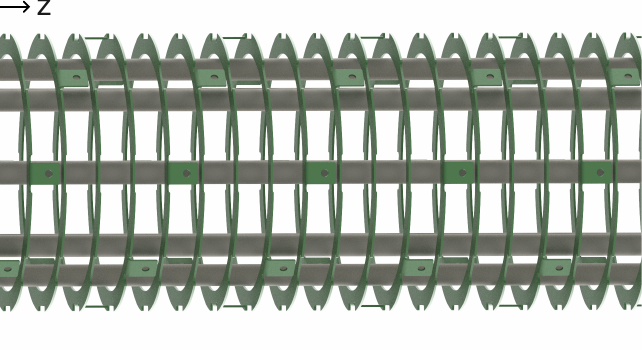}}
\caption{Schematic of the trap. The dc potentials are applied to the various rings such that in the vicinity of the trap center there is a dc potential $\Phi_z$ approximately as indicated by Eq.~\eqref{eq:lrf_potential}. For ion detection, ejection voltages are pulsed onto the rings to accelerate the ions axially out of the trapping region and onto an imaging ion detector, not shown. The rods provide radial confinement, as well as the rotating polarizing electric field required for the eEDM measurement. For more details on the electrode constructions, see Refs.~\cite{ng2023thesis, Noah_thesis, PSY_thesis}.}
\label{fig:setup}
\end{figure}

The schematic of our ion trap is shown in Fig.~\ref{fig:setup}. 

Our experiment begins when a neutral beam of ThF arrives at our ion trap in a supersonic expansion with Ne. The neutral ThF molecules are ionized as they arrive at the center of the trap via resonant-enhanced multiphoton ionization (REMPI), using techniques schemes developed in Refs~\cite{Loh_2011, Loh_2012, Grau_2012, Gresh_2016, Zhou_2019}. At the end of an experiment, we apply an axial kickout field to accelerate the ions towards a microchannel plate (MCP) detector. The ions travel for about 170~$\mu$s before they reach the MCP. The time-of-flight profile of the ions is detected by the capacitive readout of the MCP signal, and the radial profile of the ions is detected by a phosphor screen, using the techniques developed in Ref.~\cite{SHAGAM2020111257}.

Unlike standard lrf traps, which typically use four electrodes to generate a quadrupole field, our trap utilizes an eight-rod geometry. This design is required to generate the highly uniform, rotating polarizing bias electric field necessary for our precision measurement. Each rod pair, comprised of diametrically opposed rods, is driven separately giving us full control of both amplitude and phase between rod pairs. For the rf radial confinement, the two rods in a rod pair are in phase, whereas for the polarizing field, the two rods are out of phase by $\pi$. The RF signals are generated by a home-built DDS system which produces eight phase-controlled outputs. These signals are then amplified and passed through resonant drive circuits. More details on the RF signal drivers are provided in  Refs~\cite{ng2023thesis, Noah_thesis}. The typical RF drive frequency is $\omega_\mathrm{rf}/2\pi = 50$~kHz with peak-to-peak amplitude of approximately $30$~V.

\subsection{Ion Dynamics in the rrf Trap} \label{Iondynamics}
\begin{figure}[h]
\centering
\includegraphics[width=0.75\linewidth]{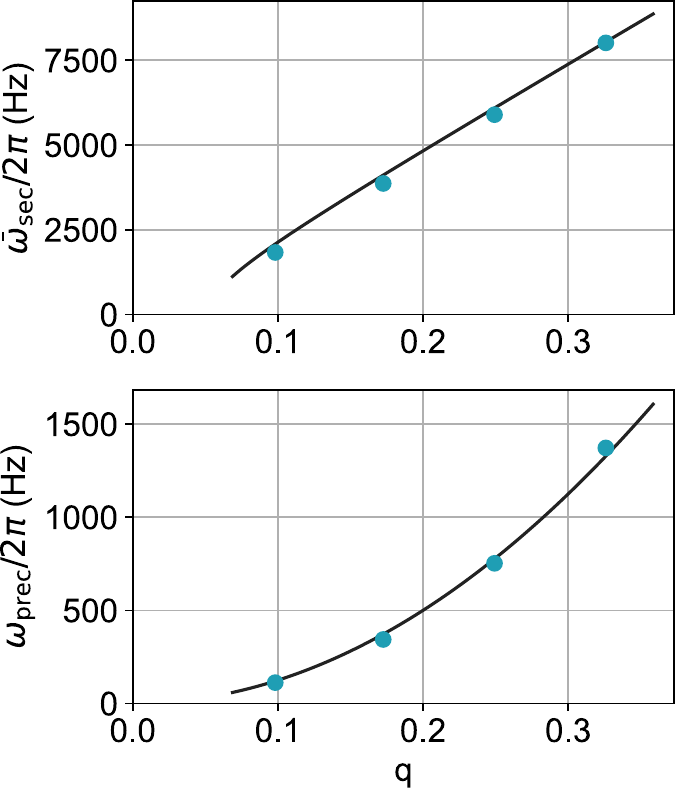}
\caption{Transverse mode frequencies in the rrf trap as a function of Mathieu parameter $q$. Blue circles show the measured mean secular frequency $\bar{\omega}_\mathrm{sec}$ (top) and precession frequency $\omega_\mathrm{prec}$ (bottom), extracted from the transverse motion following a weak kick along $x$-axis. Error bars are smaller than the marker size for all points and are not visible at this scale. Solid curves are the predictions of Eq.~\eqref{eq:rrf_freq} evaluated using the calculated trap parameters $q$ and $d$ with no fit parameters. Over the range shown, $\omega_\mathrm{prec}$ exhibits a quadratic dependence on $q$ (up to higher-order corrections).}
\label{fig:freqs}
\end{figure}

In the harmonic regime, ions in the rrf trap undergo secular motion with the two transverse eigenfrequencies given in Eq.~\eqref{eq:rrf_freq}. In our experiment, ions are excited with a relatively weak transverse electric field pulse along the $x$-axis (or $y$-axis), which prepares a linear superposition of the two circular eigenmodes with the resulting motion well within the harmonic region of the trapping potentials. The resulting transverse motion of the ions can be expressed as an oscillation at the mean secular frequency $\bar{\omega}_\mathrm{sec} = (\omega_+ + \omega_-)/2$, with a slow precession at the difference frequency ${\omega}_\mathrm{prec} = (\omega_+ - \omega_-)/2$, as previously discussed. Figure~\ref{fig:freqs} shows that both $\bar{\omega}_\mathrm{sec}$ and $\omega_\mathrm{prec}$ vary with $q$ in excellent agreement with Eq.~\eqref{eq:rrf_freq}. Equivalently, the transverse motion of ions in the rrf trap can be interpreted as that in an lrf trap in the presence of an effective axial magnetic field \cite{Bollinger_2005}. For a precession frequency of 200\,Hz, the effective magnetic field is about 60\,G for our ThF$^+$ ions. We mention this analogy as a guide to intuition, but we note the changes in ion dynamics arising from an axial magnetic field, or from the presence of rotation in the rf trapping field, are in many experiments much suppressed by the existence of static potentials breaking the degeneracy of the $\hat{x}$ and $\hat{y}$ linear motion.

\begin{figure}[h]
\centering
\includegraphics[width=0.9\linewidth]{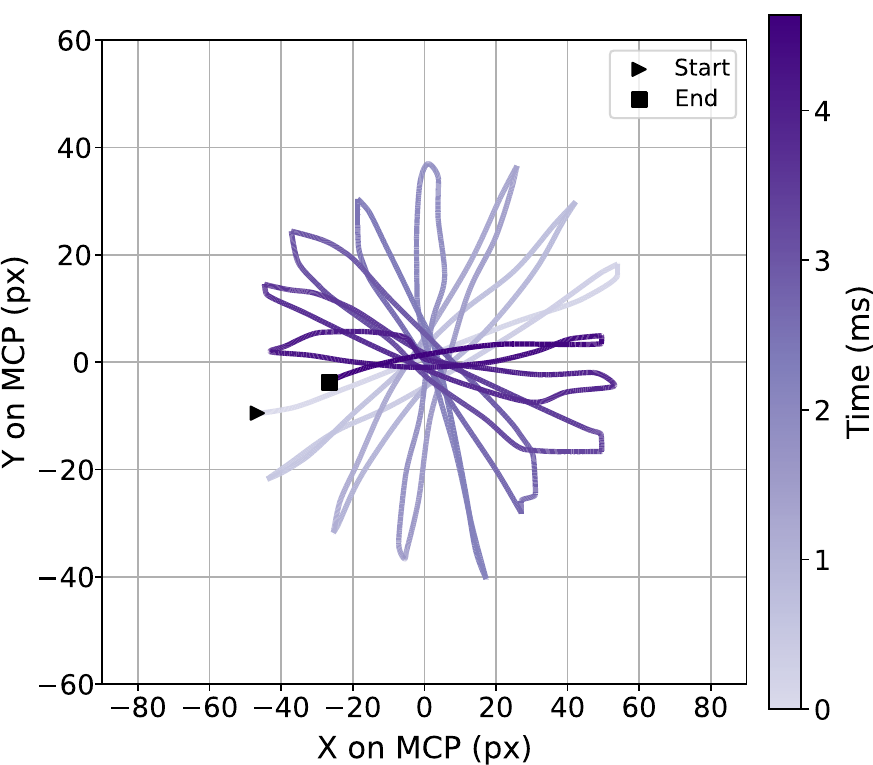}
\caption{Transverse ion trajectory in the $xy$-plane, starting approximately 4~ms after a weak transverse kick along the $x$-axis. The color darkens with increasing time delay between the kick and detection. For visualization, the data are linearly interpolated between successive points and smoothed with a moving average of two points. The trajectory shows a clear precession motion, a distinctive feature of ion dynamics in the rrf trap. The ion optics used to focus ejected ions onto an imaging detector have magnification 1.5 times larger in the $\hat{x}$ than in the $\hat{y}$ direction. We have correspondingly stretched the $x$-axis of the plot so that the image on the page is more faithful representation of the trajectory in the $xy$-plane. }
\label{fig:ion_dynamics}
\end{figure}

Figure~\ref{fig:ion_dynamics} shows the transverse ion trajectory in the rrf trap. As before, the ions are given a weak transverse electric field pulse along the $x$-axis. The plotted trajectory begins approximately 4~ms after the transverse kick, and the color darkens with increasing time between the kick and detection. For improved visualization, the data are linearly interpolated between successive points and smoothed with a 2-point moving average. The precession of the transverse motion is clearly visible. The precession frequency is expected to be $\omega_\mathrm{prec}/2\pi \approx 120$~Hz with the secular frequency $\bar{\omega}_\mathrm{sec}/2\pi\approx2000$~Hz.  Just before ejecting the ions for imaging, we quasi-adiabatically ramp down the rf amplitude, which suppresses the micromotion to the point where it is not visible in Fig.~\ref{fig:ion_dynamics}. The $\sim$$17$-petal-flower pattern in Fig.~\ref{fig:ion_dynamics} is what one would expect from the beating between simultaneously excited clockwise and counterclockwise eigenmodes with frequencies that differ by about $12\%$, and quite different from the alternating linear and circular motion one observes in beating between $\hat{x}$ and $\hat{y}$ eigenmodes.

\section{Implications for Precision Measurement in Non-ideal Geometries}
\subsection{Pseudopotential in Non-Ideal Trap Geometry }
Our ion trap, as shown in Fig.~\ref{fig:setup}, was carefully designed to provide a rotating polarizing bias electric field required for the eEDM measurement with 0.01\% uniformity. However, it was \emph{not} optimized to give perfectly harmonic rf confinement. To operate this 8-rod structure in a traditional lrf quadrupole configuration, the rods are driven with a phase pattern of $\{0, 0, \pi, \pi, 0, 0, \pi, \pi\}$. Because adjacent rods must be grouped in phase to approximate a four-pole geometry, this inherently produces 
a trap which is cylindrically symmetric for only small displacements. The anharmonic terms are highly anisotropic, characterized by a weak axis between adjacent electrodes held at the same instantaneous potential. When the trap confinement is loosened to minimize ion--ion interactions during precision spectroscopy, our ion cloud expands and samples regions where the fields deviate from the perfectly harmonic, idealized rf potentials discussed in Section~\ref{sec:idealized_rf}. 
Instead of using the idealized rf potential, one can use a pseudopotential approximation to compute an effective potential. This approximation is widely used in rf trapping and is valid when the drive frequency is large compared to the secular frequencies and the micromotion amplitude is small on the spatial scale over which the field varies \cite{GaponovMiller1958,Major2005ChargedParticleTraps,Leibfried2003}. A convenient way of expressing the rf drive is \[ \mathbf{E}(\mathbf{r},t) = \Re\!\left[\mathbf{E}_0(\mathbf{r})e^{i\omega_\mathrm{rf} t}\right], \] where $\mathbf{E}_0(\mathbf{r})$ can be a complex spatial amplitude. In the pseudopotential approximation, the time-averaged micromotion leads to a time-independent effective potential that is proportional to the square of the rf field amplitude~\cite{GaponovMiller1958,Paul1990,Major2005ChargedParticleTraps}
\begin{equation} \begin{split} 
    U_\mathrm{pseudo}(\mathbf{r}) &= \frac{{q_0}^2}{4 m_0 {\omega_\mathrm{rf}}^2} \mathbf{E(\mathbf{r}})\cdot\mathbf{E}_0(\mathbf{r})^* = \frac{{q_0}^2}{4 m_0 {\omega_\mathrm{rf}}^2} |\mathbf{E}_0(\mathbf{r})|^2 \\
    &= \frac{{q_0}^2}{4 m_0 {\omega_\mathrm{rf}}^2} |\nabla\Phi_0(\mathbf
    r)|^2,
\end{split}
\end{equation}
where $\mathbf{E}_0=-\nabla\Phi_0$ and $\Phi(\mathbf{r},t)=\Re\!\left[\Phi_0(\mathbf{r})e^{i \omega_\mathrm{rf}t}\right]$. 
% \cite{GaponovMiller1958ZhETF}
With the rf drive $\mathbf{E}(\mathbf{r},t)$ written in a complex form, it is clear that the pseudopotential approximation applies not only to the lrf potential, but to the rrf potential where the difference between them is the rrf potential having a complex amplitude $\Phi_0(\mathbf{r})$ with azimuthally varying phase. More specifically, the rrf potential of Eq.~\eqref{eq:rrf_potential} can be written as 
\[ \Phi_{\mathrm{rrf}}(r, \theta, t) \propto r^2\cos{(2\theta+\omega_\mathrm{rf} t)} = \Re\!\left[r^2 e^{i2\theta} e^{i \omega_\mathrm{rf}t}\right]. \] With $\Phi_0(r,\theta) \propto r^2 e^{i2\theta}$, the corresponding pseudopotential is 
\begin{equation} \label{eq:rrf_ponderomotive}
\begin{split}
     U_\mathrm{pseudo}(\mathbf{r}) \propto |\nabla\Phi_0|^2 = (|\nabla\Phi_c|^2 + |\nabla\Phi_s|^2),
\end{split}
\end{equation}
where $\Phi_c \propto r^2 \cos{(2\theta)}$ and $\Phi_s \propto r^2 \sin{(2\theta)}$. This decomposition provides a useful way to interpret the rrf pseudopotential discussed below.

\begin{figure*}[!hthp]
\centering
\includegraphics[width=0.99\linewidth]{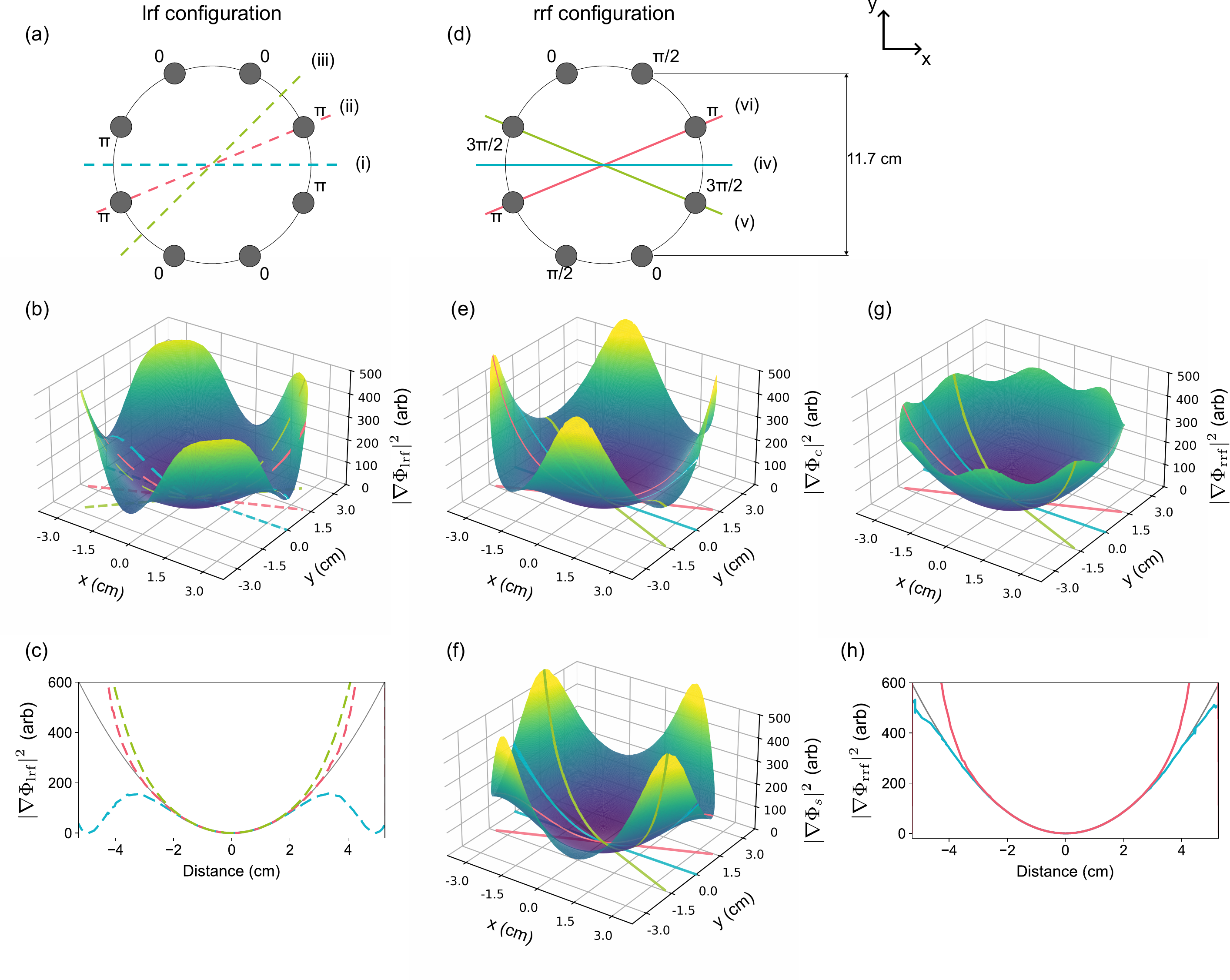}
\caption{Transverse field curvature, proportional to the pseudopotential $U_\mathrm{pseudo} \propto |\nabla\Phi|^2$, of the lrf and rrf trap configurations. (a)~Electrode phase patterns and the three geometric cuts, labeled (i)--(iii), for the lrf configuration. (b)~Surface plot of $|\nabla\Phi_\mathrm{lrf}|^2$ in the transverse plane. (c)~Corresponding one-dimensional cuts along (i)--(iii). The grey curve shows a quadratic reference fit near the trap center. (d)~Electrode phase pattern and the three geometric cuts, labeled (iv)--(vi), for the rrf configuration. (e)~and~(f) Surface plots of $|\nabla\Phi_c|^2$ and $|\nabla\Phi_s|^2$, respectively. (g)~Average of the two quadratures as the total rrf field curvature $|\nabla\Phi_\mathrm{rrf}|^2   = (|\nabla\Phi_c|^2 + |\nabla\Phi_s|^2)/2$. (h) Corresponding one-dimensional cuts along (iv)--(vi). Note that the shallowest cut in this trap is deeper than that in the lrf trap.}
\label{fig:trap_depth_cuts}
\end{figure*}

Using the pseudopotential approximation, the relevant measure of confinement and effective trap depth is the quantity $|\nabla\Phi_0|^2$. We now evaluate this quantity for our eight-rod geometry used in the experiment.
Experimentally, we drive our eight radial electrodes with fixed phase offsets. For the lrf configuration, the electrode phases are \{${0, 0, \pi, \pi, 0, 0, \pi, \pi}$\}, while for the rrf configuration they are \{${0, \pi/2, \pi, 3\pi/2, 0, \pi/2, \pi, 3\pi/2}$\}, as shown in Fig.~\ref{fig:trap_depth_cuts}a and \ref{fig:trap_depth_cuts}d, respectively. Figure~\ref{fig:trap_depth_cuts}b shows the transverse 
profile of $|\nabla\Phi_\mathrm{lrf}|^2$, a quantity to which the pseudopotential is proportional, for the lrf configuration. The effective trap depth is not azimuthally uniform. This anisotropy is seen more directly in Fig.~\ref{fig:trap_depth_cuts}c, which shows $|\nabla\Phi_0|^2$ traces along the three cuts (i), (ii), and (iii), corresponding to angles of $0^\circ, 22.5^\circ$, and $45^\circ$ with respect to the $x$-axis. The effective trap depth is shallowest along cut (i), which passes between the two electrodes held at the same instantaneous potential. Cut (ii) passes through the physical rods, spanning from $r=\pm 5.215$~cm to $r=\pm 6.485$~cm, while cut (iii) lies midway between electrodes of opposite phase.

For the rrf configuration, as discussed above, the total pseudopotential can be decomposed into the cosine and sine quadratures, shown in Fig.~\ref{fig:trap_depth_cuts}e and Fig.~\ref{fig:trap_depth_cuts}f, respectively. These two quadratures are related in space by 
\[ \Phi_c(r,\theta) = \Phi_s(r, \theta+\pi/4). \]
It is therefore tempting to think of the rrf pseudopotential as a simple azimuthal average of a single quadrature. However, the relevant quantity to the total pseudopotential is Eq.~\eqref{eq:rrf_ponderomotive}, 
the average between the two quadratures, and is shown in Fig.~\ref{fig:trap_depth_cuts}g. Figure~\ref{fig:trap_depth_cuts}h shows the total rrf $|\nabla\Phi_\mathrm{rrf}|^2$ traces along the cuts. Because of the averaging over the two quadratures, the cuts at $\pm22.5^\circ$ are identical, leaving only two unique traces. These traces are substantially more uniform than those of the lrf configuration in Fig.~\ref{fig:trap_depth_cuts}c. More importantly, the shallowest direction in the rrf trap is deeper overall than in the lrf trap.
% For our electrode geometry, this averaging of the two quadratures provides an overall trap depth that is deeper compared to the lrf configuration and, importantly, makes the trap depth more uniform as a function of azimuthal angle.
\begin{figure}[h] 
    \centering
    \includegraphics[width=0.83\linewidth]{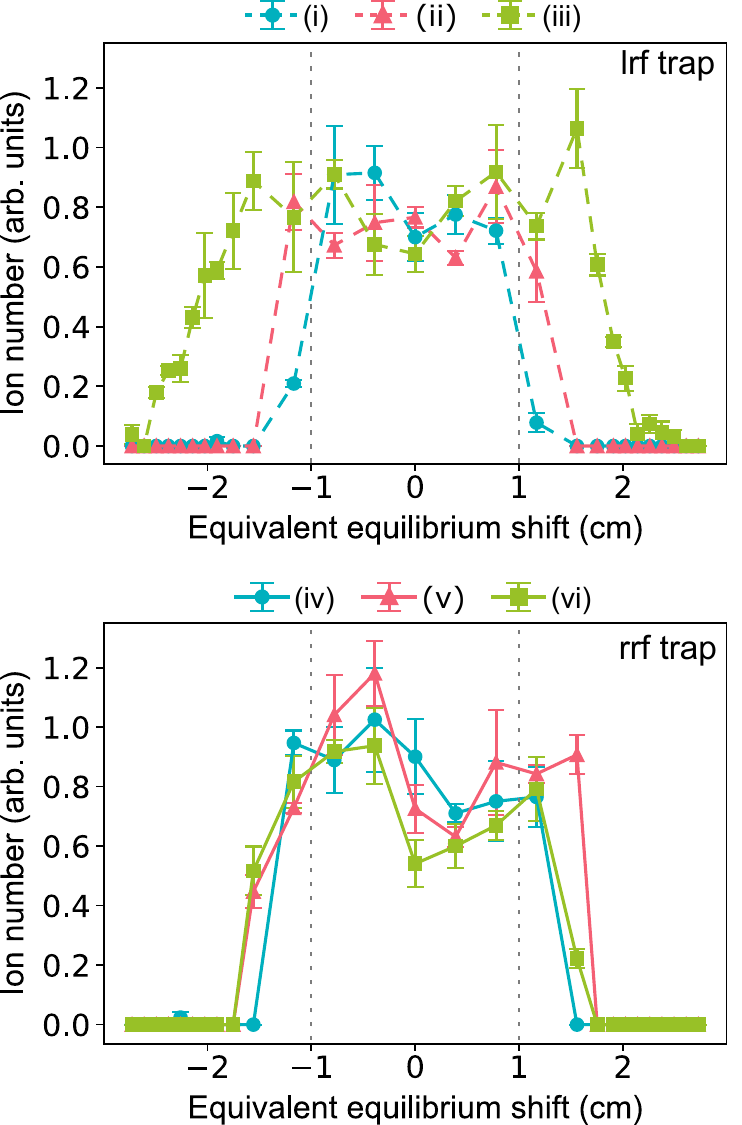}
\caption{Ion survival as a function of adiabatic dc tilt for different transverse cuts. The ion signal is recorded after the trap has been adiabatically tilted by a static dipole field. In both panels, cuts (i)-(vi) correspond to the transverse directions defined in Fig.~\ref{fig:trap_depth_cuts} (with 0$^\circ$ along the $x$-axis). The horizontal axis is the applied tilt expressed as an equivalent equilibrium shift $x_\mathrm{eq} = q_0 E_\mathrm{dc}/(m \bar{\omega}_\mathrm{sec}^2)$, defined using the harmonic approximation. The dotted vertical lines are guides to the eye. Top: cuts (i), (ii), and (iii) of the lrf configuration correspond to azimuthal angles 
0$^\circ$, 22.5$^\circ$, and 45$^\circ$, respectively. Ion loss occurs at smaller $|x_{eq}|$ for cut (i), consistent with this direction having the shallowest effective confinement. Bottom: cuts (iv), (v), and (vi) of the rrf configuration correspond to 0$^\circ$, 22.5$^\circ$, and -22.5$^\circ$, respectively. The loss curves occur at more similar $|x_{eq}|$ across angles, indicating a more azimuthally uniform effective trap depth in the rrf configuration.}
    \label{fig:trap_size}
\end{figure}

\subsection{Enhancement of Effective Trap Depth}
To experimentally probe the azimuthal dependence of the trap depth predicted in Fig.~\ref{fig:trap_depth_cuts}, we adiabatically apply a static dipole field that adds an additional potential \[ U_\mathrm{dc}(\mathbf{r}) = -q_0\mathbf{E}_\mathrm{dc}\cdot\mathbf{r}, \] which tilts the potential landscape. As $|\mathbf{E}_\mathrm{dc}|$ increases, ions are displaced from the trap center and eventually escape, so by looking at how many ions survived as a function of tilt direction, we measure the direction dependence of the effective trap depth. Here, we express the strength of the applied tilting field by the magnitude of the offset in the trap equilibrium location if the potential were fully harmonic, $x_\mathrm{eq} \propto q_0 E_\mathrm{dc}/(m \bar{\omega}_\mathrm{sec}^2)$. 

In Fig.~\ref{fig:trap_size}, in the lrf configuration, ion loss occurs at different $x_\mathrm{eq}$ for different azimuthal angles (0$^\circ$, 22.5$^\circ$, and 45$^\circ$ corresponding to cuts (i)--(iii) in the left column of Fig.~\ref{fig:trap_depth_cuts}), indicating a strong angular dependence of the effective trap depth. In particular, ions are lost at smaller tilts along the weakest direction, cut (i), consistent with the reduced trap depth for that cut predicted in Fig.~\ref{fig:trap_depth_cuts}. 

By contrast, the rrf configuration effectively reduces this asymmetry down to just two distinct axes. To explicitly test this, we measured the rrf ion survival at azimuthal angles of $0^\circ$ and $\pm 22.5^\circ$ (corresponding to cuts (iv)--(vi) in the right column of Fig.~\ref{fig:trap_depth_cuts}). Because the quadrature averaging maps the $+22.5^\circ$ and $-22.5^\circ$ directions onto the exact same averaged pseudopotential profile, the experimental loss curves for these two angles are very similar. The loss occurs at similar $x_\mathrm{eq}$ across all measured angles, visually highlighting how the rrf configuration symmetrizes the trap depth and verifies the theoretical reduction to only two distinct effective axes. Crucially, not only does ion loss occur at more uniform $x_\mathrm{eq}$ values in the rrf configuration, but the absolute minimum $x_\mathrm{eq}$ required for ions to escape is noticeably larger than that of the weakest axis in the lrf configuration.

This increased minimum escape barrier is significant because the overall effective trap depth is determined by the shallowest available axis. In our ion cloud, Coulomb interactions redistribute energy and couple their translational degrees of freedom, leading to ions exploring a wide range of trajectories over time. As a result, long-time trap depth is governed by the lowest available escape barrier of the effective potential, not by a single trajectory. The anisotropy of the lrf configuration therefore introduces ``easy escape" directions that reduce this minimum barrier and yield a comparatively shallow effective trap depth. By contrast, the rrf configuration raises this minimum barrier through quadrature averaging, resulting in a deeper overall trap depth. 

\section{Discussion}
Our ultimate goal is to maximize sensitivity to the eEDM measurement, which is proportional to $\sqrt{N}\tau$ as mentioned. The coherence time is determined largely by ion-ion collisions which depend in turn on the number of ions, their temperature, and the nature of the confining potential. We note that for holding ions in a loose yet deep potential well, there are options other than the rrf trap, including driving our eight rods in hexadecapolar pattern, for a potential more nearly approaching a square well. We are in the process of investigating which approach is most promising for maximum eEDM sensitivity. To date all our efforts to understand systematic error in our measurement of the eEDM has been in the context of the lrf ion traps \cite{Caldwell_PRA2023}. In Section~\ref{Iondynamics}, we describe the secular motion of an ion in an rrf trap being equivalent to the secular motion of an ion in an lrf trap with the addition of an effective axial magnetic field of some tens of Gauss, a value at least 1000 times larger than the physical magnetic fields present at the center of our trap during our precision measurement efforts. Of course the fictitious field couples to the center-of-mass motion of an ion and does not directly couple to the spins of the electrons within the ion. The direction of the rotation of the rrf potential and the resulting effective magnetic field do however change sign with the time reversal operation; thus we intend to explore the implications very thoroughly, given that our overarching science goal is to search for evidence for a tiny fundamental violation of time reversal symmetry. 

\section{Acknowledgment}
This work was supported by the Gordon and Betty Moore Foundation, grant DOI 10.37807/GBMF13862, by the Alfred P. Sloan Foundation under Grant No. APSF 2025-79211, by AFOSR (FA9550-20-1-0323), by NSF PFC (PHY-2317149), by the National Institute of Standards and Technology, and by a Marsico Chair of Excellence. K.~B.~Ng acknowledges support from the Banting Postdoctoral Fellowship BPF-198564, T.~A.~N. acknowledges support from the Hertz Foundation and from the NSF GRFP (DGE 2040434).


\begin{thebibliography}{33}%
\makeatletter
\providecommand \@ifxundefined [1]{%
 \@ifx{#1\undefined}
}%
\providecommand \@ifnum [1]{%
 \ifnum #1\expandafter \@firstoftwo
 \else \expandafter \@secondoftwo
 \fi
}%
\providecommand \@ifx [1]{%
 \ifx #1\expandafter \@firstoftwo
 \else \expandafter \@secondoftwo
 \fi
}%
\providecommand \natexlab [1]{#1}%
\providecommand \enquote  [1]{``#1''}%
\providecommand \bibnamefont  [1]{#1}%
\providecommand \bibfnamefont [1]{#1}%
\providecommand \citenamefont [1]{#1}%
\providecommand \href@noop [0]{\@secondoftwo}%
\providecommand \href [0]{\begingroup \@sanitize@url \@href}%
\providecommand \@href[1]{\@@startlink{#1}\@@href}%
\providecommand \@@href[1]{\endgroup#1\@@endlink}%
\providecommand \@sanitize@url [0]{\catcode `\\12\catcode `\$12\catcode `\&12\catcode `\#12\catcode `\^12\catcode `\_12\catcode `\%12\relax}%
\providecommand \@@startlink[1]{}%
\providecommand \@@endlink[0]{}%
\providecommand \url  [0]{\begingroup\@sanitize@url \@url }%
\providecommand \@url [1]{\endgroup\@href {#1}{\urlprefix }}%
\providecommand \urlprefix  [0]{URL }%
\providecommand \Eprint [0]{\href }%
\providecommand \doibase [0]{https://doi.org/}%
\providecommand \selectlanguage [0]{\@gobble}%
\providecommand \bibinfo  [0]{\@secondoftwo}%
\providecommand \bibfield  [0]{\@secondoftwo}%
\providecommand \translation [1]{[#1]}%
\providecommand \BibitemOpen [0]{}%
\providecommand \bibitemStop [0]{}%
\providecommand \bibitemNoStop [0]{.\EOS\space}%
\providecommand \EOS [0]{\spacefactor3000\relax}%
\providecommand \BibitemShut  [1]{\csname bibitem#1\endcsname}%
\let\auto@bib@innerbib\@empty
%</preamble>
\bibitem [{\citenamefont {Paul}\ and\ \citenamefont {Steinwedel}(1953)}]{Paul1953}%
  \BibitemOpen
  \bibfield  {author} {\bibinfo {author} {\bibfnamefont {W.}~\bibnamefont {Paul}}\ and\ \bibinfo {author} {\bibfnamefont {H.}~\bibnamefont {Steinwedel}},\ }\bibfield  {title} {\bibinfo {title} {Notizen: Ein neues massenspektrometer ohne magnetfeld},\ }\href {https://ui.adsabs.harvard.edu/abs/1953ZNatA...8..448P/abstract} {\bibfield  {journal} {\bibinfo  {journal} {Zeitschrift f{\"u}r Naturforschung A}\ }\textbf {\bibinfo {volume} {8}},\ \bibinfo {pages} {448} (\bibinfo {year} {1953})},\ \bibinfo {note} {{German}; often cited in English as ``A new mass spectrometer without a magnetic field''}\BibitemShut {NoStop}%
\bibitem [{\citenamefont {Cirac}\ and\ \citenamefont {Zoller}(1995)}]{CiracZoller1995}%
  \BibitemOpen
  \bibfield  {author} {\bibinfo {author} {\bibfnamefont {J.~I.}\ \bibnamefont {Cirac}}\ and\ \bibinfo {author} {\bibfnamefont {P.}~\bibnamefont {Zoller}},\ }\bibfield  {title} {\bibinfo {title} {Quantum computations with cold trapped ions},\ }\href {https://doi.org/10.1103/PhysRevLett.74.4091} {\bibfield  {journal} {\bibinfo  {journal} {Physical Review Letters}\ }\textbf {\bibinfo {volume} {74}},\ \bibinfo {pages} {4091} (\bibinfo {year} {1995})}\BibitemShut {NoStop}%
\bibitem [{\citenamefont {Monroe}\ \emph {et~al.}(1995)\citenamefont {Monroe}, \citenamefont {Meekhof}, \citenamefont {King}, \citenamefont {Itano},\ and\ \citenamefont {Wineland}}]{Monroe1995}%
  \BibitemOpen
  \bibfield  {author} {\bibinfo {author} {\bibfnamefont {C.}~\bibnamefont {Monroe}}, \bibinfo {author} {\bibfnamefont {D.~M.}\ \bibnamefont {Meekhof}}, \bibinfo {author} {\bibfnamefont {B.~E.}\ \bibnamefont {King}}, \bibinfo {author} {\bibfnamefont {W.~M.}\ \bibnamefont {Itano}},\ and\ \bibinfo {author} {\bibfnamefont {D.~J.}\ \bibnamefont {Wineland}},\ }\bibfield  {title} {\bibinfo {title} {Demonstration of a fundamental quantum logic gate},\ }\href {https://doi.org/10.1103/PhysRevLett.75.4714} {\bibfield  {journal} {\bibinfo  {journal} {Physical Review Letters}\ }\textbf {\bibinfo {volume} {75}},\ \bibinfo {pages} {4714} (\bibinfo {year} {1995})}\BibitemShut {NoStop}%
\bibitem [{\citenamefont {H{\"a}ffner}\ \emph {et~al.}(2008)\citenamefont {H{\"a}ffner}, \citenamefont {Roos},\ and\ \citenamefont {Blatt}}]{Haffner2008}%
  \BibitemOpen
  \bibfield  {author} {\bibinfo {author} {\bibfnamefont {H.}~\bibnamefont {H{\"a}ffner}}, \bibinfo {author} {\bibfnamefont {C.~F.}\ \bibnamefont {Roos}},\ and\ \bibinfo {author} {\bibfnamefont {R.}~\bibnamefont {Blatt}},\ }\bibfield  {title} {\bibinfo {title} {Quantum computing with trapped ions},\ }\href {https://doi.org/10.1016/j.physrep.2008.09.003} {\bibfield  {journal} {\bibinfo  {journal} {Physics Reports}\ }\textbf {\bibinfo {volume} {469}},\ \bibinfo {pages} {155} (\bibinfo {year} {2008})}\BibitemShut {NoStop}%
\bibitem [{\citenamefont {Blatt}\ and\ \citenamefont {Wineland}(2008)}]{BlattWineland2008}%
  \BibitemOpen
  \bibfield  {author} {\bibinfo {author} {\bibfnamefont {R.}~\bibnamefont {Blatt}}\ and\ \bibinfo {author} {\bibfnamefont {D.}~\bibnamefont {Wineland}},\ }\bibfield  {title} {\bibinfo {title} {Entangled states of trapped atomic ions},\ }\href {https://doi.org/10.1038/nature07125} {\bibfield  {journal} {\bibinfo  {journal} {Nature}\ }\textbf {\bibinfo {volume} {453}},\ \bibinfo {pages} {1008} (\bibinfo {year} {2008})}\BibitemShut {NoStop}%
\bibitem [{\citenamefont {Main}\ \emph {et~al.}(2025)\citenamefont {Main}, \citenamefont {Drmota}, \citenamefont {Nadlinger}, \citenamefont {Ainley}, \citenamefont {Agrawal}, \citenamefont {Nichol}, \citenamefont {Srinivas}, \citenamefont {Araneda},\ and\ \citenamefont {Lucas}}]{Main2025DistributedDQC}%
  \BibitemOpen
  \bibfield  {author} {\bibinfo {author} {\bibfnamefont {D.}~\bibnamefont {Main}}, \bibinfo {author} {\bibfnamefont {P.}~\bibnamefont {Drmota}}, \bibinfo {author} {\bibfnamefont {D.~P.}\ \bibnamefont {Nadlinger}}, \bibinfo {author} {\bibfnamefont {E.~M.}\ \bibnamefont {Ainley}}, \bibinfo {author} {\bibfnamefont {A.}~\bibnamefont {Agrawal}}, \bibinfo {author} {\bibfnamefont {B.~C.}\ \bibnamefont {Nichol}}, \bibinfo {author} {\bibfnamefont {R.}~\bibnamefont {Srinivas}}, \bibinfo {author} {\bibfnamefont {G.}~\bibnamefont {Araneda}},\ and\ \bibinfo {author} {\bibfnamefont {D.~M.}\ \bibnamefont {Lucas}},\ }\bibfield  {title} {\bibinfo {title} {Distributed quantum computing across an optical network link},\ }\href {https://doi.org/10.1038/s41586-024-08404-x} {\bibfield  {journal} {\bibinfo  {journal} {Nature}\ }\textbf {\bibinfo {volume} {638}},\ \bibinfo {pages} {383} (\bibinfo {year} {2025})}\BibitemShut {NoStop}%
\bibitem [{\citenamefont {Liu}\ \emph {et~al.}(2025)\citenamefont {Liu}, \citenamefont {Shaydulin}, \citenamefont {Niroula}, \citenamefont {DeCross}, \citenamefont {Hung}, \citenamefont {Kon}, \citenamefont {Cervero-Mart{\'i}n}, \citenamefont {Chakraborty}, \citenamefont {Amer}, \citenamefont {Aaronson}, \citenamefont {Acharya}, \citenamefont {Alexeev}, \citenamefont {Berg}, \citenamefont {Chakrabarti}, \citenamefont {Curchod}, \citenamefont {Dreiling}, \citenamefont {Erickson}, \citenamefont {Foltz}, \citenamefont {Foss-Feig}, \citenamefont {Hayes}, \citenamefont {Humble}, \citenamefont {Kumar}, \citenamefont {Larson}, \citenamefont {Lykov}, \citenamefont {Mills}, \citenamefont {Moses}, \citenamefont {Neyenhuis}, \citenamefont {Eloul}, \citenamefont {Siegfried}, \citenamefont {Walker}, \citenamefont {Lim},\ and\ \citenamefont {Pistoia}}]{Liu2025CertifiedRandomness}%
  \BibitemOpen
  \bibfield  {author} {\bibinfo {author} {\bibfnamefont {M.}~\bibnamefont {Liu}}, \bibinfo {author} {\bibfnamefont {R.}~\bibnamefont {Shaydulin}}, \bibinfo {author} {\bibfnamefont {P.}~\bibnamefont {Niroula}}, \bibinfo {author} {\bibfnamefont {M.}~\bibnamefont {DeCross}}, \bibinfo {author} {\bibfnamefont {S.-H.}\ \bibnamefont {Hung}}, \bibinfo {author} {\bibfnamefont {W.~Y.}\ \bibnamefont {Kon}}, \bibinfo {author} {\bibfnamefont {E.}~\bibnamefont {Cervero-Mart{\'i}n}}, \bibinfo {author} {\bibfnamefont {K.}~\bibnamefont {Chakraborty}}, \bibinfo {author} {\bibfnamefont {O.}~\bibnamefont {Amer}}, \bibinfo {author} {\bibfnamefont {S.}~\bibnamefont {Aaronson}}, \bibinfo {author} {\bibfnamefont {A.}~\bibnamefont {Acharya}}, \bibinfo {author} {\bibfnamefont {Y.}~\bibnamefont {Alexeev}}, \bibinfo {author} {\bibfnamefont {K.~J.}\ \bibnamefont {Berg}}, \bibinfo {author} {\bibfnamefont {S.}~\bibnamefont {Chakrabarti}}, \bibinfo {author} {\bibfnamefont {F.~J.}\ \bibnamefont {Curchod}}, \bibinfo {author} {\bibfnamefont
  {J.~M.}\ \bibnamefont {Dreiling}}, \bibinfo {author} {\bibfnamefont {N.}~\bibnamefont {Erickson}}, \bibinfo {author} {\bibfnamefont {C.}~\bibnamefont {Foltz}}, \bibinfo {author} {\bibfnamefont {M.}~\bibnamefont {Foss-Feig}}, \bibinfo {author} {\bibfnamefont {D.}~\bibnamefont {Hayes}}, \bibinfo {author} {\bibfnamefont {T.~S.}\ \bibnamefont {Humble}}, \bibinfo {author} {\bibfnamefont {N.}~\bibnamefont {Kumar}}, \bibinfo {author} {\bibfnamefont {J.}~\bibnamefont {Larson}}, \bibinfo {author} {\bibfnamefont {D.}~\bibnamefont {Lykov}}, \bibinfo {author} {\bibfnamefont {M.}~\bibnamefont {Mills}}, \bibinfo {author} {\bibfnamefont {S.~A.}\ \bibnamefont {Moses}}, \bibinfo {author} {\bibfnamefont {B.}~\bibnamefont {Neyenhuis}}, \bibinfo {author} {\bibfnamefont {S.}~\bibnamefont {Eloul}}, \bibinfo {author} {\bibfnamefont {P.}~\bibnamefont {Siegfried}}, \bibinfo {author} {\bibfnamefont {J.}~\bibnamefont {Walker}}, \bibinfo {author} {\bibfnamefont {C.}~\bibnamefont {Lim}},\ and\ \bibinfo {author} {\bibfnamefont
  {M.}~\bibnamefont {Pistoia}},\ }\bibfield  {title} {\bibinfo {title} {Certified randomness using a trapped-ion quantum processor},\ }\href {https://doi.org/10.1038/s41586-025-08737-1} {\bibfield  {journal} {\bibinfo  {journal} {Nature}\ }\textbf {\bibinfo {volume} {640}},\ \bibinfo {pages} {343} (\bibinfo {year} {2025})}\BibitemShut {NoStop}%
\bibitem [{\citenamefont {Diddams}\ \emph {et~al.}(2001)\citenamefont {Diddams}, \citenamefont {Udem}, \citenamefont {Bergquist} \emph {et~al.}}]{Diddams2001}%
  \BibitemOpen
  \bibfield  {author} {\bibinfo {author} {\bibfnamefont {S.~A.}\ \bibnamefont {Diddams}}, \bibinfo {author} {\bibfnamefont {T.}~\bibnamefont {Udem}}, \bibinfo {author} {\bibfnamefont {J.~C.}\ \bibnamefont {Bergquist}}, \emph {et~al.},\ }\bibfield  {title} {\bibinfo {title} {An optical clock based on a single trapped $^{199}${Hg}$^+$ ion},\ }\href {https://doi.org/10.1126/science.1061171} {\bibfield  {journal} {\bibinfo  {journal} {Science}\ }\textbf {\bibinfo {volume} {293}},\ \bibinfo {pages} {825} (\bibinfo {year} {2001})}\BibitemShut {NoStop}%
\bibitem [{\citenamefont {Ludlow}\ \emph {et~al.}(2015)\citenamefont {Ludlow}, \citenamefont {Boyd}, \citenamefont {Ye}, \citenamefont {Peik},\ and\ \citenamefont {Schmidt}}]{Ludlow2015}%
  \BibitemOpen
  \bibfield  {author} {\bibinfo {author} {\bibfnamefont {A.~D.}\ \bibnamefont {Ludlow}}, \bibinfo {author} {\bibfnamefont {M.~M.}\ \bibnamefont {Boyd}}, \bibinfo {author} {\bibfnamefont {J.}~\bibnamefont {Ye}}, \bibinfo {author} {\bibfnamefont {E.}~\bibnamefont {Peik}},\ and\ \bibinfo {author} {\bibfnamefont {P.~O.}\ \bibnamefont {Schmidt}},\ }\bibfield  {title} {\bibinfo {title} {Optical atomic clocks},\ }\href {https://doi.org/10.1103/RevModPhys.87.637} {\bibfield  {journal} {\bibinfo  {journal} {Reviews of Modern Physics}\ }\textbf {\bibinfo {volume} {87}},\ \bibinfo {pages} {637} (\bibinfo {year} {2015})}\BibitemShut {NoStop}%
\bibitem [{\citenamefont {Marshall}\ \emph {et~al.}(2025)\citenamefont {Marshall}, \citenamefont {Rodriguez~Castillo}, \citenamefont {Arthur-Dworschack}, \citenamefont {Aeppli}, \citenamefont {Kim}, \citenamefont {Lee}, \citenamefont {Warfield}, \citenamefont {Hinrichs}, \citenamefont {Nardelli}, \citenamefont {Fortier}, \citenamefont {Ye}, \citenamefont {Leibrandt},\ and\ \citenamefont {Hume}}]{Marshall2025PRL_AlClock}%
  \BibitemOpen
  \bibfield  {author} {\bibinfo {author} {\bibfnamefont {M.~C.}\ \bibnamefont {Marshall}}, \bibinfo {author} {\bibfnamefont {D.~A.}\ \bibnamefont {Rodriguez~Castillo}}, \bibinfo {author} {\bibfnamefont {W.~J.}\ \bibnamefont {Arthur-Dworschack}}, \bibinfo {author} {\bibfnamefont {A.}~\bibnamefont {Aeppli}}, \bibinfo {author} {\bibfnamefont {K.}~\bibnamefont {Kim}}, \bibinfo {author} {\bibfnamefont {D.}~\bibnamefont {Lee}}, \bibinfo {author} {\bibfnamefont {W.}~\bibnamefont {Warfield}}, \bibinfo {author} {\bibfnamefont {J.}~\bibnamefont {Hinrichs}}, \bibinfo {author} {\bibfnamefont {N.~V.}\ \bibnamefont {Nardelli}}, \bibinfo {author} {\bibfnamefont {T.~M.}\ \bibnamefont {Fortier}}, \bibinfo {author} {\bibfnamefont {J.}~\bibnamefont {Ye}}, \bibinfo {author} {\bibfnamefont {D.~R.}\ \bibnamefont {Leibrandt}},\ and\ \bibinfo {author} {\bibfnamefont {D.~B.}\ \bibnamefont {Hume}},\ }\bibfield  {title} {\bibinfo {title} {High-stability single-ion clock with $5.5\times10^{-19}$ systematic uncertainty},\ }\href
  {https://doi.org/10.1103/hb3c-dk28} {\bibfield  {journal} {\bibinfo  {journal} {Physical Review Letters}\ }\textbf {\bibinfo {volume} {135}},\ \bibinfo {pages} {033201} (\bibinfo {year} {2025})}\BibitemShut {NoStop}%
\bibitem [{\citenamefont {Lindvall}\ \emph {et~al.}(2025)\citenamefont {Lindvall}, \citenamefont {Fordell}, \citenamefont {Hanhij{\"a}rvi}, \citenamefont {Dole{\v z}al}, \citenamefont {Rahm}, \citenamefont {Weyers},\ and\ \citenamefont {Wallin}}]{Lindvall2025PRApplied_SrClock}%
  \BibitemOpen
  \bibfield  {author} {\bibinfo {author} {\bibfnamefont {T.}~\bibnamefont {Lindvall}}, \bibinfo {author} {\bibfnamefont {T.}~\bibnamefont {Fordell}}, \bibinfo {author} {\bibfnamefont {K.~J.}\ \bibnamefont {Hanhij{\"a}rvi}}, \bibinfo {author} {\bibfnamefont {M.}~\bibnamefont {Dole{\v z}al}}, \bibinfo {author} {\bibfnamefont {J.~M.}\ \bibnamefont {Rahm}}, \bibinfo {author} {\bibfnamefont {S.~L.}\ \bibnamefont {Weyers}},\ and\ \bibinfo {author} {\bibfnamefont {A.~E.}\ \bibnamefont {Wallin}},\ }\bibfield  {title} {\bibinfo {title} {$^{88}${Sr}$^{+}$ optical clock with $7.9\times10^{-19}$ systematic uncertainty and measurement of its absolute frequency with $9.8\times10^{-17}$ uncertainty},\ }\href {https://doi.org/10.1103/czlf-bfvp} {\bibfield  {journal} {\bibinfo  {journal} {Physical Review Applied}\ }\textbf {\bibinfo {volume} {24}},\ \bibinfo {pages} {044082} (\bibinfo {year} {2025})}\BibitemShut {NoStop}%
\bibitem [{\citenamefont {Cheung}\ \emph {et~al.}(2025)\citenamefont {Cheung}, \citenamefont {Porsev}, \citenamefont {Filin}, \citenamefont {Safronova}, \citenamefont {Wehrheim}, \citenamefont {Spie{\ss}}, \citenamefont {Chen}, \citenamefont {Wilzewski}, \citenamefont {Crespo L{\'o}pez-Urrutia},\ and\ \citenamefont {Schmidt}}]{Cheung2025PRL_Ni12Clock}%
  \BibitemOpen
  \bibfield  {author} {\bibinfo {author} {\bibfnamefont {C.}~\bibnamefont {Cheung}}, \bibinfo {author} {\bibfnamefont {S.~G.}\ \bibnamefont {Porsev}}, \bibinfo {author} {\bibfnamefont {D.}~\bibnamefont {Filin}}, \bibinfo {author} {\bibfnamefont {M.~S.}\ \bibnamefont {Safronova}}, \bibinfo {author} {\bibfnamefont {M.}~\bibnamefont {Wehrheim}}, \bibinfo {author} {\bibfnamefont {L.~J.}\ \bibnamefont {Spie{\ss}}}, \bibinfo {author} {\bibfnamefont {S.}~\bibnamefont {Chen}}, \bibinfo {author} {\bibfnamefont {A.}~\bibnamefont {Wilzewski}}, \bibinfo {author} {\bibfnamefont {J.~R.}\ \bibnamefont {Crespo L{\'o}pez-Urrutia}},\ and\ \bibinfo {author} {\bibfnamefont {P.~O.}\ \bibnamefont {Schmidt}},\ }\bibfield  {title} {\bibinfo {title} {Finding the {ultranarrow} $^3p_2 \rightarrow {}^3p_0$ {electric quadrupole} transition in {Ni}$^{12+}$ ion for an optical clock},\ }\href {https://doi.org/10.1103/flwf-c2m1} {\bibfield  {journal} {\bibinfo  {journal} {Physical Review Letters}\ }\textbf {\bibinfo {volume} {135}},\
  \bibinfo {pages} {093002} (\bibinfo {year} {2025})}\BibitemShut {NoStop}%
\bibitem [{\citenamefont {Willitsch}(2012)}]{Willitsch2012}%
  \BibitemOpen
  \bibfield  {author} {\bibinfo {author} {\bibfnamefont {S.}~\bibnamefont {Willitsch}},\ }\bibfield  {title} {\bibinfo {title} {{Coulomb}-crystallised molecular ions in traps: methods, applications, prospects},\ }\href {https://doi.org/10.1080/0144235X.2012.667221} {\bibfield  {journal} {\bibinfo  {journal} {International Reviews in Physical Chemistry}\ }\textbf {\bibinfo {volume} {31}},\ \bibinfo {pages} {175} (\bibinfo {year} {2012})}\BibitemShut {NoStop}%
\bibitem [{\citenamefont {Heazlewood}(2019)}]{Heazlewood2019}%
  \BibitemOpen
  \bibfield  {author} {\bibinfo {author} {\bibfnamefont {B.~R.}\ \bibnamefont {Heazlewood}},\ }\bibfield  {title} {\bibinfo {title} {Cold ion chemistry within coulomb crystals},\ }\href {https://doi.org/10.1080/00268976.2018.1564850} {\bibfield  {journal} {\bibinfo  {journal} {Molecular Physics}\ }\textbf {\bibinfo {volume} {117}},\ \bibinfo {pages} {1934} (\bibinfo {year} {2019})}\BibitemShut {NoStop}%
\bibitem [{\citenamefont {Krohn}\ and\ \citenamefont {Lewandowski}(2024)}]{KrohnLewandowski2024}%
  \BibitemOpen
  \bibfield  {author} {\bibinfo {author} {\bibfnamefont {O.~A.}\ \bibnamefont {Krohn}}\ and\ \bibinfo {author} {\bibfnamefont {H.~J.}\ \bibnamefont {Lewandowski}},\ }\bibfield  {title} {\bibinfo {title} {Cold ion--molecule reactions in the extreme environment of a {Coulomb} crystal},\ }\href {https://doi.org/10.1021/acs.jpca.3c07546} {\bibfield  {journal} {\bibinfo  {journal} {The Journal of Physical Chemistry A}\ }\textbf {\bibinfo {volume} {128}},\ \bibinfo {pages} {1737} (\bibinfo {year} {2024})}\BibitemShut {NoStop}%
\bibitem [{\citenamefont {{Harvard Natural Sciences Lecture Demonstrations}}()}]{HarvardNatSciDemosRotatingSaddleVideo}%
  \BibitemOpen
  \bibfield  {author} {\bibinfo {author} {\bibnamefont {{Harvard Natural Sciences Lecture Demonstrations}}},\ }\href {https://sciencedemonstrations.fas.harvard.edu/presentations/rotating-saddle} {\bibinfo {title} {Rotating saddle}},\ \bibinfo {howpublished} {Video},\ \bibinfo {note} {accessed 2026-02-11}\BibitemShut {NoStop}%
\bibitem [{\citenamefont {Hasegawa}\ and\ \citenamefont {Bollinger}(2005)}]{Bollinger_2005}%
  \BibitemOpen
  \bibfield  {author} {\bibinfo {author} {\bibfnamefont {T.}~\bibnamefont {Hasegawa}}\ and\ \bibinfo {author} {\bibfnamefont {J.~J.}\ \bibnamefont {Bollinger}},\ }\bibfield  {title} {\bibinfo {title} {Rotating-radio-frequency ion traps},\ }\href {https://doi.org/10.1103/PhysRevA.72.043403} {\bibfield  {journal} {\bibinfo  {journal} {Phys. Rev. A}\ }\textbf {\bibinfo {volume} {72}},\ \bibinfo {pages} {043403} (\bibinfo {year} {2005})}\BibitemShut {NoStop}%
\bibitem [{\citenamefont {Roussy}\ \emph {et~al.}(2023)\citenamefont {Roussy}, \citenamefont {Caldwell}, \citenamefont {Wright}, \citenamefont {Cairncross}, \citenamefont {Shagam}, \citenamefont {Ng}, \citenamefont {Schlossberger}, \citenamefont {Park}, \citenamefont {Wang}, \citenamefont {Ye},\ and\ \citenamefont {Cornell}}]{Roussy_Science}%
  \BibitemOpen
  \bibfield  {author} {\bibinfo {author} {\bibfnamefont {T.}~\bibnamefont {Roussy}}, \bibinfo {author} {\bibfnamefont {L.}~\bibnamefont {Caldwell}}, \bibinfo {author} {\bibfnamefont {T.}~\bibnamefont {Wright}}, \bibinfo {author} {\bibfnamefont {W.}~\bibnamefont {Cairncross}}, \bibinfo {author} {\bibfnamefont {Y.}~\bibnamefont {Shagam}}, \bibinfo {author} {\bibfnamefont {K.~B.}\ \bibnamefont {Ng}}, \bibinfo {author} {\bibfnamefont {N.}~\bibnamefont {Schlossberger}}, \bibinfo {author} {\bibfnamefont {S.~Y.}\ \bibnamefont {Park}}, \bibinfo {author} {\bibfnamefont {A.}~\bibnamefont {Wang}}, \bibinfo {author} {\bibfnamefont {J.}~\bibnamefont {Ye}},\ and\ \bibinfo {author} {\bibfnamefont {E.~A.}\ \bibnamefont {Cornell}},\ }\bibfield  {title} {\bibinfo {title} {An improved bound on the {electron's electric dipole moment}},\ }\href {https://doi.org/10.1126/science.adg4084} {\bibfield  {journal} {\bibinfo  {journal} {Science}\ }\textbf {\bibinfo {volume} {381}},\ \bibinfo {pages} {46} (\bibinfo {year}
  {2023})}\BibitemShut {NoStop}%
\bibitem [{\citenamefont {Ng}(2023)}]{ng2023thesis}%
  \BibitemOpen
  \bibfield  {author} {\bibinfo {author} {\bibfnamefont {K.~B.}\ \bibnamefont {Ng}},\ }\emph {\bibinfo {title} {The {ThF$^+$} {eEDM} experiment: concept, design, and characterization}},\ \href@noop {} {Ph.D. thesis},\ \bibinfo  {school} {University of Colorado, Boulder} (\bibinfo {year} {2023})\BibitemShut {NoStop}%
\bibitem [{\citenamefont {Cairncross}\ and\ \citenamefont {Ye}(2019)}]{Cairncross2019}%
  \BibitemOpen
  \bibfield  {author} {\bibinfo {author} {\bibfnamefont {W.~B.}\ \bibnamefont {Cairncross}}\ and\ \bibinfo {author} {\bibfnamefont {J.}~\bibnamefont {Ye}},\ }\bibfield  {title} {\bibinfo {title} {Atoms and molecules in the search for time-reversal symmetry violation},\ }\href {https://doi.org/10.1038/s42254-019-0080-0} {\bibfield  {journal} {\bibinfo  {journal} {Nature Reviews Physics}\ }\textbf {\bibinfo {volume} {1}},\ \bibinfo {pages} {510} (\bibinfo {year} {2019})}\BibitemShut {NoStop}%
\bibitem [{\citenamefont {Leibfried}\ \emph {et~al.}(2003)\citenamefont {Leibfried}, \citenamefont {Blatt}, \citenamefont {Monroe},\ and\ \citenamefont {Wineland}}]{Leibfried2003}%
  \BibitemOpen
  \bibfield  {author} {\bibinfo {author} {\bibfnamefont {D.}~\bibnamefont {Leibfried}}, \bibinfo {author} {\bibfnamefont {R.}~\bibnamefont {Blatt}}, \bibinfo {author} {\bibfnamefont {C.}~\bibnamefont {Monroe}},\ and\ \bibinfo {author} {\bibfnamefont {D.}~\bibnamefont {Wineland}},\ }\bibfield  {title} {\bibinfo {title} {Quantum dynamics of single trapped ions},\ }\href {https://doi.org/10.1103/RevModPhys.75.281} {\bibfield  {journal} {\bibinfo  {journal} {Reviews of Modern Physics}\ }\textbf {\bibinfo {volume} {75}},\ \bibinfo {pages} {281} (\bibinfo {year} {2003})}\BibitemShut {NoStop}%
\bibitem [{\citenamefont {Major}\ \emph {et~al.}(2005)\citenamefont {Major}, \citenamefont {Gheorghe},\ and\ \citenamefont {Werth}}]{Major2005ChargedParticleTraps}%
  \BibitemOpen
  \bibfield  {author} {\bibinfo {author} {\bibfnamefont {F.~G.}\ \bibnamefont {Major}}, \bibinfo {author} {\bibfnamefont {V.~N.}\ \bibnamefont {Gheorghe}},\ and\ \bibinfo {author} {\bibfnamefont {G.}~\bibnamefont {Werth}},\ }\href {https://doi.org/10.1007/b137836} {\emph {\bibinfo {title} {Charged Particle Traps: Physics and Techniques of Charged Particle Field Confinement}}}\ (\bibinfo  {publisher} {Springer},\ \bibinfo {address} {Berlin, Heidelberg},\ \bibinfo {year} {2005})\BibitemShut {NoStop}%
\bibitem [{\citenamefont {Schlossberger}(2023)}]{Noah_thesis}%
  \BibitemOpen
  \bibfield  {author} {\bibinfo {author} {\bibfnamefont {N.}~\bibnamefont {Schlossberger}},\ }\emph {\bibinfo {title} {An Apparatus for Measuring the Electron's Electric Dipole Moment in Trapped {ThF$^+$}}},\ \href {https://scholar.colorado.edu/concern/graduate_thesis_or_dissertations/kk91fn485} {\bibinfo {type} {{Ph.D.} thesis}},\ \bibinfo  {school} {University of Colorado Boulder}, \bibinfo {address} {Boulder, Colorado} (\bibinfo {year} {2023})\BibitemShut {NoStop}%
\bibitem [{\citenamefont {Park}(2026)}]{PSY_thesis}%
  \BibitemOpen
  \bibfield  {author} {\bibinfo {author} {\bibfnamefont {S.~Y.}\ \bibnamefont {Park}},\ }\emph {\bibinfo {title} {Achieving long coherence time for a multiplexed, ion-based eEDM experiment}},\ \href@noop {} {\bibinfo {type} {Doctoral thesis}},\ \bibinfo  {school} {University of Colorado Boulder}, \bibinfo {address} {Boulder, CO} (\bibinfo {year} {2026})\BibitemShut {NoStop}%
\bibitem [{\citenamefont {Loh}\ \emph {et~al.}(2011)\citenamefont {Loh}, \citenamefont {Wang}, \citenamefont {Grau}, \citenamefont {Yahn}, \citenamefont {Field}, \citenamefont {Greene},\ and\ \citenamefont {Cornell}}]{Loh_2011}%
  \BibitemOpen
  \bibfield  {author} {\bibinfo {author} {\bibfnamefont {H.}~\bibnamefont {Loh}}, \bibinfo {author} {\bibfnamefont {J.}~\bibnamefont {Wang}}, \bibinfo {author} {\bibfnamefont {M.}~\bibnamefont {Grau}}, \bibinfo {author} {\bibfnamefont {T.~S.}\ \bibnamefont {Yahn}}, \bibinfo {author} {\bibfnamefont {R.~W.}\ \bibnamefont {Field}}, \bibinfo {author} {\bibfnamefont {C.~H.}\ \bibnamefont {Greene}},\ and\ \bibinfo {author} {\bibfnamefont {E.~A.}\ \bibnamefont {Cornell}},\ }\bibfield  {title} {\bibinfo {title} {Laser-induced fluorescence studies of {HfF$^+$} produced by autoionization},\ }\href {https://doi.org/10.1063/1.3652333} {\bibfield  {journal} {\bibinfo  {journal} {The Journal of Chemical Physics}\ }\textbf {\bibinfo {volume} {135}},\ \bibinfo {pages} {154308} (\bibinfo {year} {2011})}\BibitemShut {NoStop}%
\bibitem [{\citenamefont {Loh}\ \emph {et~al.}(2012)\citenamefont {Loh}, \citenamefont {Stutz}, \citenamefont {Yahn}, \citenamefont {Looser}, \citenamefont {Field},\ and\ \citenamefont {Cornell}}]{Loh_2012}%
  \BibitemOpen
  \bibfield  {author} {\bibinfo {author} {\bibfnamefont {H.}~\bibnamefont {Loh}}, \bibinfo {author} {\bibfnamefont {R.~P.}\ \bibnamefont {Stutz}}, \bibinfo {author} {\bibfnamefont {T.~S.}\ \bibnamefont {Yahn}}, \bibinfo {author} {\bibfnamefont {H.}~\bibnamefont {Looser}}, \bibinfo {author} {\bibfnamefont {R.~W.}\ \bibnamefont {Field}},\ and\ \bibinfo {author} {\bibfnamefont {E.~A.}\ \bibnamefont {Cornell}},\ }\bibfield  {title} {\bibinfo {title} {{REMPI} spectroscopy of {HfF}},\ }\href {https://doi.org/10.1016/j.jms.2012.06.014} {\bibfield  {journal} {\bibinfo  {journal} {Journal of Molecular Spectroscopy}\ }\textbf {\bibinfo {volume} {276--277}},\ \bibinfo {pages} {49} (\bibinfo {year} {2012})}\BibitemShut {NoStop}%
\bibitem [{\citenamefont {Grau}\ \emph {et~al.}(2012)\citenamefont {Grau}, \citenamefont {Leanhardt}, \citenamefont {Loh}, \citenamefont {Sinclair}, \citenamefont {Stutz}, \citenamefont {Yahn},\ and\ \citenamefont {Cornell}}]{Grau_2012}%
  \BibitemOpen
  \bibfield  {author} {\bibinfo {author} {\bibfnamefont {M.}~\bibnamefont {Grau}}, \bibinfo {author} {\bibfnamefont {A.~E.}\ \bibnamefont {Leanhardt}}, \bibinfo {author} {\bibfnamefont {H.}~\bibnamefont {Loh}}, \bibinfo {author} {\bibfnamefont {L.~C.}\ \bibnamefont {Sinclair}}, \bibinfo {author} {\bibfnamefont {R.~P.}\ \bibnamefont {Stutz}}, \bibinfo {author} {\bibfnamefont {T.~S.}\ \bibnamefont {Yahn}},\ and\ \bibinfo {author} {\bibfnamefont {E.~A.}\ \bibnamefont {Cornell}},\ }\bibfield  {title} {\bibinfo {title} {Near-infrared {LIF} spectroscopy of {HfF}},\ }\href {https://doi.org/10.1016/j.jms.2011.12.006} {\bibfield  {journal} {\bibinfo  {journal} {Journal of Molecular Spectroscopy}\ }\textbf {\bibinfo {volume} {272}},\ \bibinfo {pages} {32} (\bibinfo {year} {2012})}\BibitemShut {NoStop}%
\bibitem [{\citenamefont {Gresh}\ \emph {et~al.}(2016)\citenamefont {Gresh}, \citenamefont {Cossel}, \citenamefont {Zhou}, \citenamefont {Ye},\ and\ \citenamefont {Cornell}}]{Gresh_2016}%
  \BibitemOpen
  \bibfield  {author} {\bibinfo {author} {\bibfnamefont {D.~N.}\ \bibnamefont {Gresh}}, \bibinfo {author} {\bibfnamefont {K.~C.}\ \bibnamefont {Cossel}}, \bibinfo {author} {\bibfnamefont {Y.}~\bibnamefont {Zhou}}, \bibinfo {author} {\bibfnamefont {J.}~\bibnamefont {Ye}},\ and\ \bibinfo {author} {\bibfnamefont {E.~A.}\ \bibnamefont {Cornell}},\ }\bibfield  {title} {\bibinfo {title} {Broadband velocity modulation spectroscopy of {ThF$^+$} for use in a measurement of the electron electric dipole moment},\ }\href {https://doi.org/10.1016/j.jms.2015.11.001} {\bibfield  {journal} {\bibinfo  {journal} {Journal of Molecular Spectroscopy}\ }\textbf {\bibinfo {volume} {319}},\ \bibinfo {pages} {1} (\bibinfo {year} {2016})}\BibitemShut {NoStop}%
\bibitem [{\citenamefont {Zhou}\ \emph {et~al.}(2019)\citenamefont {Zhou}, \citenamefont {Ng}, \citenamefont {Cheng}, \citenamefont {Gresh}, \citenamefont {Field}, \citenamefont {Ye},\ and\ \citenamefont {Cornell}}]{Zhou_2019}%
  \BibitemOpen
  \bibfield  {author} {\bibinfo {author} {\bibfnamefont {Y.}~\bibnamefont {Zhou}}, \bibinfo {author} {\bibfnamefont {K.~B.}\ \bibnamefont {Ng}}, \bibinfo {author} {\bibfnamefont {L.}~\bibnamefont {Cheng}}, \bibinfo {author} {\bibfnamefont {D.~N.}\ \bibnamefont {Gresh}}, \bibinfo {author} {\bibfnamefont {R.~W.}\ \bibnamefont {Field}}, \bibinfo {author} {\bibfnamefont {J.}~\bibnamefont {Ye}},\ and\ \bibinfo {author} {\bibfnamefont {E.~A.}\ \bibnamefont {Cornell}},\ }\bibfield  {title} {\bibinfo {title} {Visible and ultraviolet laser spectroscopy of {ThF}},\ }\href {https://doi.org/10.1016/j.jms.2019.02.006} {\bibfield  {journal} {\bibinfo  {journal} {Journal of Molecular Spectroscopy}\ }\textbf {\bibinfo {volume} {358}},\ \bibinfo {pages} {1} (\bibinfo {year} {2019})}\BibitemShut {NoStop}%
\bibitem [{\citenamefont {Shagam}\ \emph {et~al.}(2020)\citenamefont {Shagam}, \citenamefont {Cairncross}, \citenamefont {Roussy}, \citenamefont {Zhou}, \citenamefont {Ng}, \citenamefont {Gresh}, \citenamefont {Grogan}, \citenamefont {Ye},\ and\ \citenamefont {Cornell}}]{SHAGAM2020111257}%
  \BibitemOpen
  \bibfield  {author} {\bibinfo {author} {\bibfnamefont {Y.}~\bibnamefont {Shagam}}, \bibinfo {author} {\bibfnamefont {W.~B.}\ \bibnamefont {Cairncross}}, \bibinfo {author} {\bibfnamefont {T.~S.}\ \bibnamefont {Roussy}}, \bibinfo {author} {\bibfnamefont {Y.}~\bibnamefont {Zhou}}, \bibinfo {author} {\bibfnamefont {K.~B.}\ \bibnamefont {Ng}}, \bibinfo {author} {\bibfnamefont {D.~N.}\ \bibnamefont {Gresh}}, \bibinfo {author} {\bibfnamefont {T.}~\bibnamefont {Grogan}}, \bibinfo {author} {\bibfnamefont {J.}~\bibnamefont {Ye}},\ and\ \bibinfo {author} {\bibfnamefont {E.~A.}\ \bibnamefont {Cornell}},\ }\bibfield  {title} {\bibinfo {title} {Continuous temporal ion detection combined with time-gated imaging: Normalization over a large dynamic range},\ }\href {https://doi.org/10.1016/j.jms.2020.111257} {\bibfield  {journal} {\bibinfo  {journal} {Journal of Molecular Spectroscopy}\ }\textbf {\bibinfo {volume} {368}},\ \bibinfo {pages} {111257} (\bibinfo {year} {2020})}\BibitemShut {NoStop}%
\bibitem [{\citenamefont {Gaponov}\ and\ \citenamefont {Miller}(1958)}]{GaponovMiller1958}%
  \BibitemOpen
  \bibfield  {author} {\bibinfo {author} {\bibfnamefont {A.~V.}\ \bibnamefont {Gaponov}}\ and\ \bibinfo {author} {\bibfnamefont {M.~A.}\ \bibnamefont {Miller}},\ }\bibfield  {title} {\bibinfo {title} {Potential wells for charged particles in a high-frequency electromagnetic field},\ }\href {https://www.jetp.ras.ru/cgi-bin/dn/e_007_01_0168.pdf} {\bibfield  {journal} {\bibinfo  {journal} {Soviet Physics JETP}\ }\textbf {\bibinfo {volume} {7}},\ \bibinfo {pages} {168} (\bibinfo {year} {1958})},\ \bibinfo {note} {{Russian} original: Zh. Eksp. Teor. Fiz. 34, 242--243 (1958)}\BibitemShut {NoStop}%
\bibitem [{\citenamefont {Paul}(1990)}]{Paul1990}%
  \BibitemOpen
  \bibfield  {author} {\bibinfo {author} {\bibfnamefont {W.}~\bibnamefont {Paul}},\ }\bibfield  {title} {\bibinfo {title} {Electromagnetic traps for charged and neutral particles},\ }\href {https://doi.org/10.1103/RevModPhys.62.531} {\bibfield  {journal} {\bibinfo  {journal} {Reviews of Modern Physics}\ }\textbf {\bibinfo {volume} {62}},\ \bibinfo {pages} {531} (\bibinfo {year} {1990})}\BibitemShut {NoStop}%
\bibitem [{\citenamefont {Caldwell}\ \emph {et~al.}(2023)\citenamefont {Caldwell}, \citenamefont {Roussy}, \citenamefont {Wright}, \citenamefont {Cairncross}, \citenamefont {Shagam}, \citenamefont {Ng}, \citenamefont {Schlossberger}, \citenamefont {Park}, \citenamefont {Wang}, \citenamefont {Ye},\ and\ \citenamefont {Cornell}}]{Caldwell_PRA2023}%
  \BibitemOpen
  \bibfield  {author} {\bibinfo {author} {\bibfnamefont {L.}~\bibnamefont {Caldwell}}, \bibinfo {author} {\bibfnamefont {T.~S.}\ \bibnamefont {Roussy}}, \bibinfo {author} {\bibfnamefont {T.}~\bibnamefont {Wright}}, \bibinfo {author} {\bibfnamefont {W.~B.}\ \bibnamefont {Cairncross}}, \bibinfo {author} {\bibfnamefont {Y.}~\bibnamefont {Shagam}}, \bibinfo {author} {\bibfnamefont {K.~B.}\ \bibnamefont {Ng}}, \bibinfo {author} {\bibfnamefont {N.}~\bibnamefont {Schlossberger}}, \bibinfo {author} {\bibfnamefont {S.~Y.}\ \bibnamefont {Park}}, \bibinfo {author} {\bibfnamefont {A.}~\bibnamefont {Wang}}, \bibinfo {author} {\bibfnamefont {J.}~\bibnamefont {Ye}},\ and\ \bibinfo {author} {\bibfnamefont {E.~A.}\ \bibnamefont {Cornell}},\ }\bibfield  {title} {\bibinfo {title} {Systematic and statistical uncertainty evaluation of the {HfF}$^{+}$ {electron's electric dipole moment} experiment},\ }\href {https://doi.org/10.1103/PhysRevA.108.012804} {\bibfield  {journal} {\bibinfo  {journal} {Phys. Rev. A}\ }\textbf {\bibinfo
  {volume} {108}},\ \bibinfo {pages} {012804} (\bibinfo {year} {2023})}\BibitemShut {NoStop}%
\end{thebibliography}
\end{document}